\documentclass[a4paper, amsfonts, amssymb, amsmath, reprint, showkeys, nofootinbib, twoside, onecolumn,notitlepage]{revtex4-1}

\def\mnras{{MNRAS}}             
\def\prd{{Phys.~Rev.~D}}        
\usepackage{amsmath,amstext}
\usepackage[T1]{fontenc}
\usepackage{amssymb}
\usepackage{graphicx}
\usepackage{ae,aecompl}

\DeclareFontFamily{OT1}{pzc}{}
\DeclareFontShape{OT1}{pzc}{m}{it}{<-> s * [1.10] pzcmi7t}{}
\DeclareMathAlphabet{\mathpzc}{OT1}{pzc}{m}{it}

\usepackage{hyperref}
\usepackage{amsmath}
\usepackage{amssymb}
\usepackage{mathtools}
\usepackage{bm}
\usepackage{cleveref}
\usepackage{tensor}
\usepackage{braket}
\usepackage{enumitem}
\usepackage{mhchem}
\usepackage{amsthm}
\usepackage{nccmath}
\usepackage{mathrsfs}
\usepackage{color}

\newcommand{\spc}{\quad \quad \quad}

\newcommand{\K}{{\mathcal{K}}}

\def\be{\begin{equation}}
\def\ee{\end{equation}}
\def\beq{\begin{eqnarray}}
\def\eeq{\end{eqnarray}}

\theoremstyle{definition}

\theoremstyle{theorem}
\newtheorem{theorem}{Theorem}

\begin{document}
\title{Heat propagation in rotating relativistic bodies}
\author{L.~Gavassino$^1$ and M.~Antonelli$^2$}
\affiliation{$^1$Department of Mathematics, Vanderbilt University, Nashville, TN, USA\\
$^2$CNRS/IN2P3, Laboratoire de Physique Corpusculaire de Caen, 14050 Caen, France}

\begin{abstract}
We investigate heat propagation in rigidly rotating bodies within the theory of general relativity. Using a first-order gradient expansion, we derive a universal partial differential equation governing the temperature evolution. This equation is hyperbolic, causal, and stable, and it naturally accounts for both rotational and gravitational Tolman-Ehrenfest effects. Any other first-order theory consistent with established physics (including the parabolic theories used in neutron star cooling models) must be equivalent to our formulation within an error that is of higher order in gradients. As a case study, we analyze heat transfer in solid cylinders rotating around their symmetry axis, so that the tangential speed approaches the speed of light on the surface. We also compute the relativistic rotational corrections to the cooling law of black bodies.
\end{abstract} 

\maketitle
\section{Introduction}
\vspace{-0.4cm}

Currently, no relativistically covariant formula describing the propagation of heat in rotating bodies is known. 
Ideally, one would wish to have a \textit{single} partial differential equation for the temperature field $T(x^\alpha)$ alone, incorporating both special and general relativistic effects, such as time dilation, length contraction, relativistic causality, and the Tolman-Ehrenfest effect (i.e., the non-uniformity of the temperature in equilibrium due to gravity or rotation \cite{TolmanLaw1930,landau5,cercignani_book,RovelliSmerlak2011,Santiago:2019aem}), all framed in a manifestly covariant language. Such an equation should be ``universal''. That is, it should apply to all isotropic solids \cite[\S 32]{landau7} and cold highly-viscous fluids \cite[\S 50]{landau6} that undergo rigid rotation, irrespective of the system's detailed microphysics (provided that the mean free path is small compared to the macroscopic gradients \cite{landau6,huang_book,Struchtrup2011Review,Grozdanov:2019kge,HellerHydrohedron2023jtd,GavassinoInfiniteOrded2024pgl}). Finally, it should reduce to the usual heat equation in the non-relativistic limit:
\vspace{-0.15cm}
\begin{equation}\label{newtonian}
(\partial_t {+}v^j \partial_j)T =\dfrac{1}{nc_v} \partial_j \big(\kappa \partial^j T\big) \, ,
\end{equation}
where $v^j(x^\alpha)$ is a rigid particle flow, $n(x^\alpha)$ is the particle density, $c_v(T,n)$ is the specific heat per particle, and $\kappa(T,n)$ is the heat conductivity. Some equations were proposed \cite{ThorneStellarStructure1967,lopez2011,Lander:2018cvq,Kim:2023lta}, but none of them fulfill \textit{all} the requirements above.

To understand the difficulty of the problem, let us consider the following naive relativistic generalization of~\eqref{newtonian}:
\vspace{-0.15cm}
\begin{equation}\label{trivailcovariantize}
u^\mu \nabla_\mu T =\dfrac{1}{nc_v} \Delta^{\mu \nu} \nabla_\mu \big(\kappa \nabla_\nu T\big) \, ,   
\end{equation}
where $u^\mu(x^\alpha)$ is the particle flow velocity, and $\Delta^{\mu \nu}=g^{\mu \nu}{+}u^\mu u^\nu$ is the projector orthogonal to $u^\mu$. At first glance, the above equation appears reasonable; however, it presents several issues. Most notably, the state $T(x^\alpha)=\text{const}$ is an equilibrium solution, which contradicts the Tolman-Ehrenfest effect, according to which, in thermal equilibrium, rotating bodies exhibit higher temperatures farther from the rotation axis \cite[\S 12.5]{cercignani_book}, while self-gravitating bodies are hotter at their centers. Additionally, equation \eqref{trivailcovariantize} is parabolic \cite[\S 7.1]{evansPDEbook}, so its initial value problem is well-posed only for initial data specified on hypersurfaces everywhere orthogonal to $u^\mu$ \cite{Kost2000}. Due to Frobenius' theorem \cite[\S B.3]{Wald}, such hypersurfaces do not exist in rotating systems, rendering \eqref{trivailcovariantize} unsolvable. These ill-posedness issues stem from the acausal nature of \eqref{trivailcovariantize}, which produces instabilities in reference frames where the medium is moving~\cite{Hiscock_Insatibility_first_order,GavassinoLyapunov_2020,GavassinoSuperlum2021,GavassinoBounds2023myj}.

Given the aforementioned challenges, state-of-the-art simulations of heat transport in neutron stars \cite{Schaab:1998ed,Negreiros2012,Yakovlev:2004iq,Potekhin:2015qsa,Beznogov:2022wae} employ the following workaround, originally proposed in \cite{Miralles1993}. First, express the equations of the underlying microscopic theory (in their case, radiation kinetic theory) in a coordinate system tailored to the symmetries of the problem. Next, consider the energy conservation equation, and apply a series of simplifications and approximations, which invariably involve discarding certain time derivatives. The final result is an equation that resembles a diffusion equation in the chosen coordinates. Such an equation lacks general covariance and causality, forcing the user to stick to the selected coordinates, with no possibility to, e.g., boost (as the boosted initial value problem would be ill-posed). Furthermore, this equation is not guaranteed to be universal, since its derivation relies on a specific microscopic model. 

In this article, we finally settle this issue by deriving a causal and well-posed relativistic generalization of \eqref{newtonian}, and by showing that such generalization must be unique. Specifically, our main result is the following:
\vspace{-0.1cm}
\begin{theorem}\label{theo1}
Consider a medium whose particle flow $u^\mu(x^\alpha)$ is rigid in Born's sense, i.e. there exists a scalar function $\K(x^\alpha)\,{>}\, 0$ such that $\K u^\alpha$ is a Killing vector \textup{\cite{Born1909,Herglotz1910,Noether1910}}. Suppose that particles and rest-frame energy are conserved, i.e. $\nabla_\mu (nu^\mu)\,{=}\,0$ and $u_\nu \nabla_\mu T^{\mu \nu}\,{=}\,0$. Then, there exists only one local field equation for $T(x^\alpha)$ (up to field redefinitions), involving solely $\{u^\mu,\K,T,n\}$, which \textup{(a)} is manifestly covariant, \textup{(b)} is consistent with the Tolman-Ehrenfest effect, and \textup{(c)} does not allow for any flow of heat in equilibrium. Such an equation can be expressed as follows:
\vspace{-0.15cm}
\begin{equation}\label{BDNKHeat}
\boxed{u^\mu \nabla_\mu (\K T) =\dfrac{1}{nc_v} \nabla_\mu \big[\kappa \nabla^\mu (\K T) \big] +\mathcal{O}(\nabla^3)  \, .}
\end{equation}
\end{theorem}

We will first prove the above theorem. Then, we will discuss some interesting applications.

{\color{black}We note that the mathematical assumptions underlying Theorem \ref{theo1} implicitly rely upon physical assumptions that define (and put limitations on) the domain of applicability of \eqref{BDNKHeat}. For instance, the fact that the field equation depends only on $\{u^\mu,\K,T,n\}$ implies that local anisotropies associated with lattice structures or magnetic fields can be neglected. Moreover, the requirement that $\K u^\mu$ be a timelike Killing vector indicates that the metric can be treated as a stationary background, whose geometry is not affected by the transfers of heat.}

Throughout the article, we adopt the metric signature $(-,+,+,+)$, and work in natural units, with $c=k_B=\hbar=1$.

\section{Formal derivation}

In this section, we present the proof of Theorem~\ref{theo1}. Our reasoning follows the same logic as in \cite{Bemfica2019_conformal1,Kovtun2019,BemficaDNDefinitivo2020}, where one writes down the most general expression for conserved currents up to first order in derivatives, and then performs field redefinitions to simplify the equations as much as possible, while still keeping the equations hyperbolic.


\subsection{The relevant conservation law of the problem}\label{heatcurrentsection}

Our starting point is the rigidity assumption, according to which the vector field $\K^\mu=\K u^\mu$ obeys the Killing equation $\nabla_{(\mu} \K_{\nu)}=0$.\footnote{
    Note that requiring $\K u^\mu$ to be a Killing vector also implies that the body's rotation rate must remain constant \cite{Herglotz1910,Noether1910}. In fact, in special relativity, any change in the rotation rate leads to a variation in the ratio ``perimeter''/``radius'' (due to length contraction), thereby violating the rigidity assumption. Consequently, equation \eqref{BDNKHeat} is valid only in systems where the temperature of the medium is much lower than the mass of its constituent particles. If this condition is not met, heat exchange can significantly alter the body's moment of inertia (since ``$E = mc^2$''), enforcing a change in rotation rate to conserve angular momentum. This would violate the condition that $\K u^\mu$ is a Killing vector, thereby breaking Born rigidity. Another reason for assuming low temperature relative to mass is that it allows us to neglect the gravitational field variations induced by heat transfer. Otherwise, such variations would likely eliminate any exact spacetime symmetries, and no Killing vectors would exist.}
Combining this with energy conservation, we find that the four-current $J_H^\mu=-T^{\mu \nu}\K_\nu$ is conserved \cite[\S 3.2]{Hawking1973}, namely
\begin{equation}\label{conserviamo}
\nabla_\mu J^\mu_H =-(\nabla_\mu T^{\mu\nu}) \K u_\nu- T^{\mu \nu}\nabla_{(\mu}\K_{\nu)}=0 \, .
\end{equation}
Physically, this vector field can be interpreted as a ``redshifted heat current''. To see why, we note that, independently of the microscopic details, the stress-energy tensor can always be geometrically decomposed as \cite{Eckart40,Israel_Stewart_1979,Hishcock1983,rezzolla_book}
\begin{equation}\label{decompo}
T^{\mu \nu}=\varepsilon u^\mu u^\nu +u^\mu q^\nu + q^\mu u^\nu  +P^{\mu \nu} \, ,
\end{equation}
where $\varepsilon$ is the energy density, $q^\mu$ is the heat flux ($u_\mu q^\mu=0$), and $P^{\mu \nu}$ is the stress tensor ($u_\mu P^{\mu \nu}=P^{\mu \nu}u_\nu=0$), all viewed from the perspective of the observer comoving with the particle flow. Then, we have that
\begin{equation}\label{JHHH}
J_H^\mu =\K (\varepsilon u^\mu + q^\mu) \, .
\end{equation}
We see that, aside from the factor $\K$ (which accounts for redshift \cite[\S 27]{landau5}), the time component of the vector field $J_H^\mu$ in the particle's rest frame is the thermal energy density, and the space components form the heat flux. This clearly qualifies $J_H^\mu$ as a heat four-current (see also~\cite{Van:2007pw,Van:2008cy}).

We emphasize that the rigidity assumption is crucial for deriving a conserved energy current whose evolution is governed solely by heat transfer. In a general flow, the divergence of $\varepsilon u^\mu \,{+}\,q^\mu$ is also sensitive to the value of $P^{\mu \nu}$, which reflects the fact that thermal energy can be produced via ``$PdV$'' work or via viscous heating~\cite[\S 50]{landau6}. 

We also note that, if one splits the energy density $\varepsilon$ into the rest-mass part $mn$ and the internal part $\varepsilon_I$, then the rest-mass contribution to the heat current is independently conserved: $\nabla_\mu (mn\K^\mu)=0$. Hence, the heat current may be replaced with $J_{I}^\mu=\K(\varepsilon_I u^\mu +q^\mu)$, which facilitates the comparison with the Newtonian theory.

\subsection{Equilibrium states}

The Tolman-Ehrenfest effect states that, in global thermodynamic equilibrium, the vector field $u^\mu/T$ is a Killing
vector \cite{TolmanLaw1930,Israel1981,BecattiniBeta2016,Santiago:2019aem,GavassinoGibbs2021,GavassinoStabilityCarter2022} (see Appendix~\ref{AppAAA} for a quick derivation). On the other hand, we already know that $\K u^\mu$ is a Killing vector, by
the rigidity assumption. Hence, we conclude that, in thermodynamic equilibrium, the following holds:
\begin{equation}\label{KTconst}
\K T =\text{const} \, .
\end{equation}
Any theory of heat conduction that is consistent with the Tolman-Ehrenfest effect must admit \eqref{KTconst} as a solution. Additionally, assumption (c) of Theorem~\ref{theo1} states that, along the solution \eqref{KTconst}, the heat flux $q^\mu$ vanishes, giving
\begin{equation}
J_H^\mu \propto  u^\mu \spc (\text{whenever }\K T=\text{const}) \, .
\end{equation}

\subsection{Redefining the non-equilibrium temperature}\label{nonequo}
\vspace{-0.2cm}

Before we can prove Theorem~\ref{theo1}, we need one final ingredient: field redefinitions. Let us recall that, strictly speaking, the temperature field $T$ is uniquely defined only in equilibrium. In a non-equilibrium state, one might still attempt to define $T$ by ``inserting a thermometer'' into the medium, but different thermometers may yield slightly different readings \cite{Kovtun2019,KovtunTemperature2022vas}. This implies that the field $T(x^\alpha)$ is just an effective degree of freedom that we use to parameterize the local state of the system, and there are multiple options for how to define it out of equilibrium. Indeed, fixed some definition of $T$, any other scalar field $\Tilde{T}$ that equals $T$ in equilibrium is equally acceptable. Up to first order in gradients, a generic scalar field $\Tilde{T}$, which is a local functional of $\{u^\mu, \K,T,n \}$, can always be decomposed as follows:
\begin{equation}\label{glaudio}
\Tilde{T}=F_1(\K,T,n)+F_2(\K,T,n) \nabla_\lambda u^\lambda +F_3(\K,T,n)u^\lambda \nabla_\lambda \K +F_4(\K,T,n)u^\lambda \nabla_\lambda T+F_5(\K,T,n)u^\lambda \nabla_\lambda n+\mathcal{O}(\nabla^2)\, ,
\end{equation}
where we have just written down all possible Lorentz scalars involving only first powers of derivatives, and $F_i$ are arbitrary functions. In practice, most of the first-order terms in \eqref{glaudio} vanish, since the Killing condition on $\K u^\mu$ entails that $\nabla_\lambda u^\lambda=u^\lambda \nabla_\lambda \K=0$, and the conservation law $\nabla_\lambda (nu^\lambda)=0$ entails that $u^\lambda \nabla_\lambda n=0$. Thus, we are left with
\begin{equation}\label{glaudio2}
\Tilde{T}=F_1(\K,T,n)  +F_4(\K,T,n)u^\lambda \nabla_\lambda T+\mathcal{O}(\nabla^2)\, ,
\end{equation}
Of course, if we wish $\Tilde{T}$ to qualify as a non-equilibrium proxy for the local temperature, it should be equal to $T$ along all equilibrium states \eqref{KTconst}, which gives $F_1=T$, and it should not depend on the local value of the redshift function $\K$. Therefore, arrive at the following class of ``equally good'' notions of non-equilibrium temperature:
\begin{equation}\label{glaudio3}
\Tilde{T}=T  -\tau (T,n)u^\lambda \nabla_\lambda T+\mathcal{O}(\nabla^2)\, ,
\end{equation}
where we renamed $F_4$ as $-\tau$.

Equation \eqref{glaudio3} has a rather intuitive physical interpretation: If the temperature of the medium varies over time due to heat exchanges, the particles need some time to adjust to the new thermal conditions. As a result, the microscopic distribution deviates from a perfect (relativistic) Maxwellian, and the relative magnitude of such deviations is roughly proportional to $(\tau u^\lambda \nabla_\lambda T)/T$, for some characteristic relaxation time $\tau$. Choosing between $T$ and $\Tilde{T}$ amounts to choosing different methods of approximating a non-Maxwellian distribution with a Maxwellian fit.

\vspace{-0.2cm}
\subsection{Step-by-step proof of the Theorem}
\vspace{-0.2cm}

We are now ready to prove Theorem~\ref{theo1}. Our first observation is that, whatever the equation of motion for $T$ may be, it must be compatible with the conservation law $\nabla_\mu J_H^\mu =0$. Hence, rather than determining the differential equation for $T$ directly, we can first determine $J_H^\mu$, and then determine the field equation from its divergence. This leaves us with the task of constructing the most general expression for $J_H^\mu$ involving all possible Lorentz structures one can build out of $\{u^\mu, \K,T,n \}$. To first order in gradients, we have
\begin{equation}\label{BDNKov}
\begin{split}
J^\mu_H ={}& A_1(\K,T,n) u^\mu+A_2(\K,T,n) u^\lambda \nabla_\lambda u^\mu +A_3(\K ,T,n)u^\mu \nabla_\lambda u^\lambda \\ 
+{}& A_4 (\K,T,n) \nabla^\mu \K  + A_5(\K,T,n) u^\mu u^\lambda \nabla_\lambda \K   \\
+{}& A_6(\K,T,n)\nabla^\mu T+A_7(\K,T,n) u^\mu u^\lambda \nabla_\lambda T\\
+{}& A_8(\K,T,n)\nabla^\mu n+A_9(\K,T,n) u^\mu u^\lambda \nabla_\lambda n +\mathcal{O}(\nabla^2) \, ,\\
\end{split}    
\end{equation}
where $A_i$ are some arbitrary functions. Our goal now is to simplify the above formula as much as possible, using symmetries, field redefinitions, and first-principle constraints coming from physics. 

Our first observation is that, since $\K u^\mu$ is a Killing vector, and $nu^\mu$ is a conserved current, we can again eliminate some terms thanks to the relations $\nabla_\lambda u^\lambda\,{=}\,0$, $u^\lambda \nabla_\lambda \K\,{=}\,0$, $\K u^\lambda \nabla_\lambda u^\mu \,{=}\,\nabla^\mu \K$, and $u^\lambda \nabla_\lambda n\,{=}\,0$. Thus, we are left with
\begin{equation}\label{BDNKov2}
\begin{split}
J^\mu_H ={}& A_1(\K,T,n) u^\mu+\big[A_2(\K,T,n)/\K + A_4 (\K,T,n)\big] \nabla^\mu \K    \\
+{}& A_6(\K,T,n)\nabla^\mu T+A_7(\K,T,n) u^\mu u^\lambda \nabla_\lambda T+ A_8(\K,T,n)\nabla^\mu n +\mathcal{O}(\nabla^2) \, .\\
\end{split}    
\end{equation}
Next, we recall that the scalar field $\K$ is defined up to a multiplicative constant. In fact, if $\K u^\mu$ is a Killing vector, then also $a\K u^\mu$ with $a=\text{const}$ is a Killing vector. This defines a symmetry of the theory, according to which the transformation $\K \rightarrow a\K$ should induce a transformation $J_H^\mu \rightarrow aJ_H^\mu$ in the current (recall that $J_H^\mu=-T^{\mu \nu}\K_\nu$). This symmetry constrains the dependence of $A_i$ on $\K$, giving
\begin{equation}\label{BDNKov3}
\begin{split}
J^\mu_H ={}& \K B_1(T,n) u^\mu+B_2(T,n) \nabla^\mu \K   +\K B_3(T,n)\nabla^\mu T \\
+{}& \K B_4 (T,n) u^\mu u^\lambda \nabla_\lambda T+\K B_5(T,n)\nabla^\mu n +\mathcal{O}(\nabla^2) \, .\\
\end{split}    
\end{equation}
Now, to further simplify the equation, we can use field redefinitions of the kind discussed in section~\ref{nonequo}, i.e.  $T=\Tilde{T}+\tau(\Tilde{T},n)u^\lambda \nabla_\lambda \Tilde{T}+\mathcal{O}(\nabla^2)$. The result is
\begin{equation}\label{BDNKov4}
\begin{split}
J^\mu_H ={}& \K B_1(\Tilde{T},n) u^\mu+B_2(\Tilde{T},n) \nabla^\mu \K   +\K B_3(\Tilde{T},n)\nabla^\mu \Tilde{T} \\
+{}& \K \big[ B_4 (\Tilde{T},n)+\tau (\Tilde{T},n)\partial_T B_1(\Tilde{T},n)\big] u^\mu u^\lambda \nabla_\lambda \Tilde{T}+\K B_5(\Tilde{T},n)\nabla^\mu n +\mathcal{O}(\nabla^2) \, .\\
\end{split}    
\end{equation}
Note that the factors multiplying all gradient terms have been evaluated directly at temperature $\Tilde{T}$ instead of $T$, since the related error to $J_H^\mu$ scales like $(T{-}\Tilde{T}) \times \nabla \sim \nabla^2$. Setting $\tau=-B_4/\partial_T B_1$, the term proportional to $u^\lambda \nabla_\lambda \Tilde{T}$ in \eqref{BDNKov4} vanishes.
Therefore, dropping the ``tilde'' symbol for notational convenience, we now have
\begin{equation}\label{BDNKov5}
\begin{split}
J^\mu_H ={}& \K B_1(T,n) u^\mu+B_2(T,n) \nabla^\mu \K   +\K B_3(T,n)\nabla^\mu T +\K B_5(T,n)\nabla^\mu n +\mathcal{O}(\nabla^2) \, .\\
\end{split}    
\end{equation}
The final step is to require that, in an equilibrium state (i.e. in a state with $\K T=\text{const}$), the heat current must be proportional to $u^\mu$. This forces the coefficient $B_5$ to vanish, because the gradient $\nabla^\mu n$ is orthogonal to $u^\mu$, and it survives in states with $\K T=\text{const}$. Furthermore, we must impose that $B_2 =TB_3$, because the terms proportional to $\nabla^\mu \K$ and $\nabla^\mu T$ should combine into one single term proportional to $\nabla^\mu(\K T)$. Therefore, we arrive at the expression
\begin{equation}\label{BDNKone}
J_H^\mu =\K \varepsilon_{\text{eq}}(T,n)u^\mu -\kappa(T,n)\nabla^\mu (\K T)+\mathcal{O}(\nabla^2) \, ,
\end{equation}
where we have renamed $B_1$ as $\varepsilon_{\text{eq}}$, and $B_3$ as $-\kappa$. Imposing the conservation law $\nabla_\mu J_H^\mu =0$, and recalling that $\nabla_\mu u^\mu =u^\mu \nabla_\mu \K=u^\mu \nabla_\mu n =0$, we finally obtain the desired equation (where we define $nc_v=\partial_T \varepsilon_{\text{eq}}$):
\begin{equation}
nc_v u^\mu \nabla_\mu (\K T)-\nabla_\mu \big[\kappa \nabla^\mu (\K T) \big]+\mathcal{O}(\nabla^3) =0 \, .
\end{equation}
This equation, truncated at order $\nabla^2$, is hyperbolic and causal. In fact, its principal part is $\propto \nabla_\mu \nabla^\mu T$, which means that information propagates exactly at the speed of light, and the initial value problem is well-posed \cite[Th. 10.1.3]{Wald}. Furthermore, in the flat non-rotating case, we recover the usual telegrapher equation $\partial_t (\kappa \partial_t T){+} nc_v \partial_t T=\partial_j (\kappa \partial^j T)$ \cite[\S 6.5.1]{rezzolla_book}, which is known to be dynamically stable in all reference frames~\cite{Hishcock1983,GavassinoCausality2021,GavassinoUniveraalityI2023odx}.

\vspace{-0.2cm}
\subsection{Eckart temperature and Fourier law}
\vspace{-0.2cm}

To better understand the physical content of \eqref{BDNKHeat}, let us analyze equation \eqref{BDNKone} more closely. Comparing \eqref{BDNKone} with \eqref{JHHH}, we obtain the following first-order constitutive relations for energy density and energy flux:
\begin{equation}\label{ourframe}
\begin{split}
\varepsilon={}&\varepsilon_{\text{eq}}(T,n)+\kappa(T,n) u^\mu \nabla_\mu T +\mathcal{O}(\nabla^2) \, , \\
q^\mu ={}& -\kappa(T,n) \Delta^{\mu \nu } \dfrac{\nabla_\nu (\K T)}{\K} +\mathcal{O}(\nabla^2) \, . \\
\end{split}
\end{equation}
The first equation tells us that our definition of temperature out of equilibrium does not match the one of Eckart \cite{Eckart40}, according to which $\varepsilon\equiv \varepsilon_\text{eq}(T_{\text{Eckart}},n)$. Physically, Eckart's choice amounts to fitting the non-equilibrium distribution with a Maxwellian with the same total energy. Our choice is different, since $\varepsilon-\varepsilon_{\text{eq}}(T,n)=\mathcal{O}(\nabla)\neq 0$. If one insists on adopting Eckart's definition, they only need to make the field redefinition $T=T_{\text{Eckart}}-\frac{\kappa}{nc_v}u^\mu \nabla_\mu T_{\text{Eckart}}+\mathcal{O}(\nabla^2)$. However, the resulting equation of motion, namely
\begin{equation}\label{eckartone}
u^\mu \nabla_\mu (\K T_{\text{Eckart}})=\dfrac{1}{nc_v} \nabla_\mu \big[\kappa \Delta^{\mu \nu}\nabla_\nu (\K T_{\text{Eckart}}) \big] +\mathcal{O}(\nabla^3) \, ,
\end{equation}
turns out to be parabolic, and its initial value problem is ill-posed for rotating systems, analogously to the naive equation \eqref{trivailcovariantize}. This makes $T_{\text{Eckart}}$ a rather inconvenient choice of temperature, if the goal is to solve the equation, as opposed to, e.g., computing $q^\mu (x^\alpha)$ for a previously assigned profile $T_{\text{Eckart}}(x^\alpha)$ \cite{Geroch95}. Instead, \eqref{BDNKHeat} always admits a well-posed initial value formulation (for initial data prescribed on spacelike hypersurfaces).

Finally, let us examine the formula for the heat flux $q^\mu$ more closely. Starting from \eqref{ourframe}, and recalling that $\K u^\lambda \nabla_\lambda u_\nu =\nabla_\nu \K$, we can write the following chain of identities:
\begin{equation}\label{heattttone}
\begin{split}
q^\mu ={}& -\kappa T \Delta^{\mu \nu } \dfrac{\nabla_\nu (\K T)}{\K T} +\mathcal{O}(\nabla^2) \\
={}& -\kappa T \Delta^{\mu \nu } \bigg[ \dfrac{\nabla_\nu T}{T} +\dfrac{\nabla_\nu \K}{\K}\bigg] +\mathcal{O}(\nabla^2) \\
={}& -\kappa T \Delta^{\mu \nu }\bigg[ \dfrac{\nabla_\nu T}{T} + u^\lambda \nabla_\lambda u_\nu\bigg] +\mathcal{O}(\nabla^2) \, , \\
\end{split}    
\end{equation}
which is the relativistic Fourier law that one finds in textbooks \cite[Exercise 22.7]{MTW_book}. The acceleration term is a purely relativistic correction. It arises because the matter-frame energy $-u_\mu p^\mu$ of a mediator particle that transports heat (e.g. a photon) suffers from redshift as the particle travels between two accelerated material elements \cite[\S 2.5]{NovikovThorne1973}.

\section{Application to relativistic stars}

As a consistency check, let us verify that, when we specialize \eqref{BDNKHeat} to relativistic stars, we obtain a hyperbolic generalization of the standard equations that are used in the neutron-star literature \cite{Schaab:1998ed,Negreiros2012,Yakovlev:2004iq,Potekhin:2015qsa,Beznogov:2022wae,Miralles1993}.

\subsection{Non-rotating case}

We consider a spherically symmetric static spacetime, whose metric tensor reads
\begin{equation}
ds^2 =-e^{2\Phi} dt^2 +e^{2\Lambda} dr^2 + r^2 (d\theta^2 +\sin^2 \!\theta\, d\phi^2) \, ,
\end{equation}
with $\Phi=\Phi(r)$ and $\Lambda=\Lambda(r)$. The medium filling this spacetime is assumed at rest, meaning that $u^\mu \partial_\mu = e^{-\Phi} \partial_t$. Since $\partial_t$ is a Killing vector, this motion is indeed rigid (in the sense of Theorem~\ref{theo1}), and we have $\K =e^\Phi$. Thus, the Tolman-Ehrenfest effect requires that the redshifted temperature $e^\Phi T$ be constant in equilibrium, which is a well-known fact. Then, the relativistic heat equation \eqref{BDNKHeat} becomes
\begin{equation}\label{fshiszko}
e^{-\Phi} \partial_t (e^\Phi T) =\dfrac{1}{nc_v \sqrt{-g}} \partial_\mu \big[\kappa \sqrt{-g} g^{\mu \nu}\partial_\nu (e^\Phi T) \big] +\mathcal{O}(\partial^3) \, ,
\end{equation}
where the square root of the determinant of the metric can be decomposed as $\sqrt{-g}=e^{\Phi}\sqrt{G}$, where $\sqrt{G}=e^\Lambda r^2 \sin\theta$ is the square root of the determinant of the space part of the metric. 
This allows us to rearrange \eqref{fshiszko} as
\begin{equation}\label{fshiszko2}
\partial_t (e^\Phi T) =\dfrac{1}{nc_v \sqrt{G}} \partial_i \big[\kappa e^\Phi \sqrt{G} g^{ij}\partial_j (e^\Phi T) \big]-\dfrac{1}{nc_v} e^{-\Phi} \partial_t \big[\kappa \partial_t  (e^\Phi T) \big]+\mathcal{O}(\partial^3) \, ,
\end{equation}
which is the usual equation used in neutron-star models \cite{Yakovlev:2004iq,Potekhin:2015qsa,Lander:2018cvq,pons_vigano_review_2019}, except for an additional term with two time derivatives. Such a new term is needed for covariance, because it keeps the equations hyperbolic and causal, which guarantees that the initial value problem remains well-posed if we decide to change coordinates. On the other hand, it should be noted that, if $T$ solves \eqref{fshiszko2}, then $\partial_t (e^\Phi T) =\mathcal{O}(\partial^2)$. Hence, the quantity $\partial_t \big[\kappa \partial_t  (e^\Phi T) \big]$ really scales like a \textit{third-order} derivative term (as long as $T$ is a solution of the field equation). Therefore, if we stick to the coordinates $\{t,r,\theta,\phi \}$, we may reabsorb the last term into the error $\mathcal{O}(\partial^3)$, and we finally obtain the usual parabolic equation
\begin{equation}\label{fshiszko3}
\partial_t (e^\Phi T) =\dfrac{1}{nc_v \sqrt{G}} \partial_i \big[\kappa e^\Phi \sqrt{G} g^{ij}\partial_j (e^\Phi T) \big]+\mathcal{O}(\partial^3) \, .
\end{equation}

\subsection{Rotating case with axial symmetry}

We consider an axially symmetric stationary spacetime, with metric tensor
\begin{equation}
ds^2 =-e^{2\Phi} dt^2 +A^2 (dr^2+r^2 d\theta^2) +B^2 r^2 \sin^2 \theta (d\phi-\omega dt)^2 \, ,
\end{equation}
where $\Phi=\Phi(r,\theta)$, $A=A(r,\theta)$, $B=B(r,\theta)$, and $\omega=\omega(r,\theta)$. The medium filling this spacetime is assumed to be rigidly rotating, meaning that its four-velocity is $u^\mu \partial_\mu =e^{-\Phi} \Gamma  (\partial_t +\Omega \partial_\phi)$, with $\Omega=\text{const}$, and
\begin{equation}
\Gamma = \dfrac{1}{\sqrt{1-e^{-2\Phi}(\Omega{-}\omega)^2 B^2 r^2\sin^2\theta}}.
\end{equation}
The rigidity of this motion follows from the fact that $\partial_t+\Omega\partial_\phi$ is indeed a Killing vector, which implies that $\K=e^{\Phi}/\Gamma$. As is usually done in the literature, let us assume that also the temperature profile is axially symmetric, i.e. $\partial_\phi T=0$. Then, equation \eqref{BDNKHeat} becomes
\begin{equation}
\begin{split}
e^{-\Phi} \Gamma \,  \partial_t\bigg(\dfrac{e^\Phi T}{\Gamma}\bigg)={}& \dfrac{1}{nc_v \sqrt{-g}} \partial_r \bigg[\kappa g^{rr} \sqrt{-g} \partial_r \bigg(\dfrac{e^\Phi T}{\Gamma}\bigg) \bigg] \\
+{}& \dfrac{1}{nc_v \sqrt{-g}} \partial_\theta \bigg[\kappa g^{\theta \theta} \sqrt{-g} \partial_\theta \bigg(\dfrac{e^\Phi T}{\Gamma}\bigg) \bigg] \\
+{}& \dfrac{1}{nc_v \sqrt{-g}} \partial_t \bigg[\kappa g^{tt} \sqrt{-g} \partial_t \bigg(\dfrac{e^\Phi T}{\Gamma}\bigg) \bigg] +\mathcal{O}(\partial^3) \, ,\\
\end{split}
\end{equation}
with $\sqrt{-g}=e^\Phi A^2 B r^2 \sin\theta$. Again, this equation is hyperbolic and causal due to the second time-derivative in the last line. However, also in this case, $\partial_t(e^\Phi T/\Gamma)$ is of order $\partial^2$ along solutions of the field equation. Thus, the last line can be absorbed into $\mathcal{O}(\partial^3)$, and we arrive at the parabolic equation that is commonly used in the literature \cite{Negreiros2012,Beznogov:2022wae}:
\begin{equation}
\partial_t\bigg(\dfrac{e^\Phi T}{\Gamma}\bigg)= \dfrac{1}{nc_v \Gamma A^2 B r^2 \sin\theta}\bigg\{ \partial_r \bigg[\kappa  e^\Phi B r^2 \sin\theta \partial_r \bigg(\dfrac{e^\Phi T}{\Gamma}\bigg) \bigg] +\partial_\theta \bigg[\kappa  e^\Phi  B  \sin\theta \partial_\theta \bigg(\dfrac{e^\Phi T}{\Gamma}\bigg) \bigg] \bigg\}  +\mathcal{O}(\partial^3)\, .
\end{equation}

\section{Spinning cylinders}

We can now examine some physical implications of equation \eqref{BDNKHeat}. Let us recall that this work aims to explore how special relativity gives rise to new physical phenomena associated with heat propagation at fast rotation rates. Therefore, it will be most instructive to focus on thermal processes inside infinitely long cylinders, which rotate rigidly about their symmetry axis in Minkowski spacetime.

\subsection{Thermal evolution equation}

The natural coordinates in a cylinder are $\{t, \varrho, \phi, z\}$, corresponding respectively to time, radial distance from the cylinder's axis, angular position, and position along the axis. All these coordinates are defined relatively to an inertial observer comoving with the cylinder's axis. The Minkowski line element then reads
\begin{equation}
ds^2 =-dt^2+d\varrho^2 +\varrho^2 d\phi^2 +dz^2 \, .
\end{equation}
Rigid rotation entails that
$u^\mu \partial_\mu =\Gamma(\partial_t+\Omega\partial_\phi)$ with $\Omega=\text{const}$, since $\partial_t$ and $\partial_\phi$ are Killing vectors (and thus $\K=1/\Gamma$). The proportionality function $\Gamma$ is the usual Lorentz factor, namely $\Gamma=(1-\Omega^2\varrho^2)^{-1/2}$, which exists only for $\varrho \Omega<1$. From this, we immediately conclude that the radius of the cylinder must be finite, and it cannot exceed $\Omega^{-1}$, otherwise parts of the cylinder would revolve faster than light. 

The Tolman-Ehrenfest effect requires that $T/\Gamma$ is constant in equilibrium, which gives us the well-known formula
\begin{equation}\label{Tolmaniamo}
T_{\text{Equilibrium}} =\dfrac{T_{\text{axis}}}{\sqrt{1-\Omega^2\varrho^2}} \, ,
\end{equation}
according to which the temperature increases with distance from the rotation axis \cite[\S 12.5.a]{cercignani_book}.
This phenomenon is known as the ``inertia of heat'' \cite{Israel1981}.
Thus, applying Theorem~\ref{theo1}, we obtain the following evolution equation:
\begin{equation}
(\partial_t{+}\Omega\partial_\phi)\bigg( \dfrac{T}{\Gamma} \bigg)=\dfrac{1}{nc_v\Gamma \varrho} \partial_\mu \bigg[\kappa \varrho g^{\mu \nu}\partial_\nu \bigg(\dfrac{T}{\Gamma} \bigg) \bigg] +\mathcal{O}(\partial^3) \, .
\end{equation}
For notational convenience, let us introduce the ``redshifted'' temperature
$T_R=T/\Gamma$. Furthermore, let us assume that $\kappa$ and $nc_v$ are constant, and let us introduce the thermal diffusivity $D=\kappa/(nc_v)$. Then, we find that
\begin{equation}\label{grumpyone}
(\partial_t{+}\Omega\partial_\phi)T_R=\dfrac{D}{\Gamma } \bigg[ \dfrac{1}{\varrho} \partial_\varrho \big( \varrho \partial_\varrho T_R\big)+\dfrac{1}{\varrho^2}  \partial_\phi^2 T_R+ \partial_z^2 T_R- \partial_t^2 T_R\bigg] +\mathcal{O}(\partial^3)\, .
\end{equation}
Again, the term with two time derivatives renders the equation hyperbolic. If we stick to the given coordinate system, we may get rid of $\partial_t^2 T_R$ by noticing that, along solutions of \eqref{grumpyone}, we have the scaling $\partial_t T_R=-\Omega \partial_\phi T_R +\mathcal{O}(\partial^2)$, which implies that $\partial_t^2 T=\Omega^2 \partial^2_\phi T+\mathcal{O}(\partial^3)$, resulting in the parabolic equation (see also \cite{Basar:2024qxd,GavassinoParabolic2025hwz})
\begin{equation}\label{grainofsalt}
(\partial_t{+}\Omega\partial_\phi)T_R=\dfrac{D}{\Gamma } \bigg[ \dfrac{1}{\varrho} \partial_\varrho \big( \varrho \partial_\varrho T_R\big)+\dfrac{1 }{\varrho^2\Gamma^2}  \partial_\phi^2 T_R+ \partial_z^2 T_R\bigg] +\mathcal{O}(\partial^3) \, .
\end{equation}
As can be seen, special relativity places corrections everywhere. First, one needs to evolve $T_R=T/\Gamma$, rather than $T$, to account for the Tolman-Ehrenfest effect. Second, heat propagation in all directions is suppressed by a factor $1/\Gamma$, which accounts for time dilation. Finally, heat propagation in direction $\phi$ is suppressed by an additional factor $1/\Gamma^2$, which accounts for length contraction (twice) \cite{Basar:2024qxd}. The only effect that does not show up in \eqref{grainofsalt} is the relativity of simultaneity, which enters as a higher-order correction in the derivative expansion. This is apparent if one considers that the speed of propagation of information changed from $1$ in \eqref{grumpyone} to $\infty$ in~\eqref{grainofsalt}.

\subsection{First application: Stationary heat flux across a cylindrical shell}\label{SectionB}

Physics students in a futuristic civilization might need to solve homework problems such as the following: A cylindrical shell with uniform properties (i.e. constant $\kappa$ and $nc_v$) extends from an inner radius $\varrho_{\text{in}}$ to an outer radius $\varrho_{\text{out}}$ and is rigidly rotating at relativistic speeds. 
A thermostat located at the inner boundary enforces a temperature $T(\varrho_{\text{in}})=T_{\text{in}}$, while another thermostat at the outer boundary maintains $T(\varrho_{\text{out}})=T_{\text{out}}$. Determine the stationary temperature distribution and the resulting heat flux across the shell.

\begin{figure}[h!]
    \centering
\includegraphics[width=0.49\linewidth]{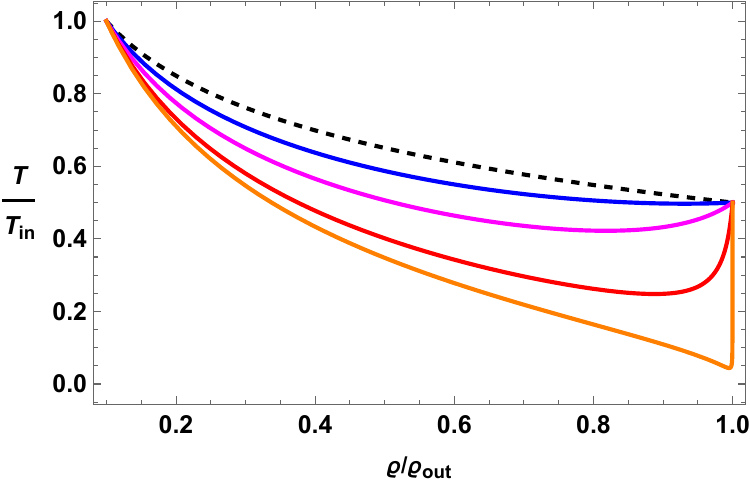}
\includegraphics[width=0.49\linewidth]{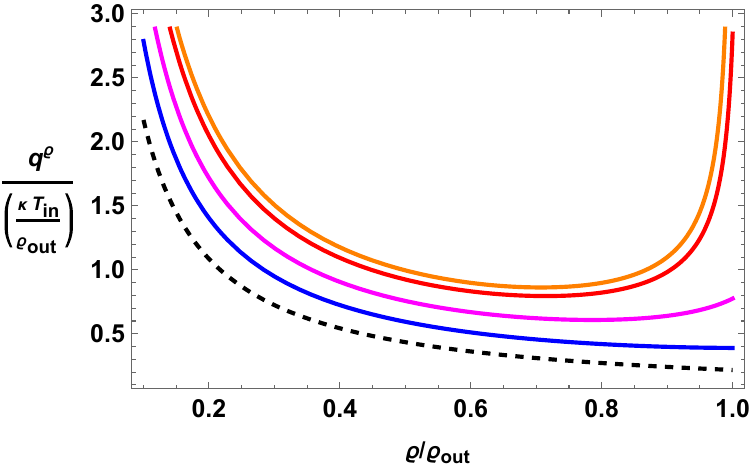}
\caption{Temperature (left panel) and heat flux (right panel) across a cylindrical shell rotating with outer tangential speed $\Omega \varrho_{\text{out}}=0$ (dashed), 0.7 (blue), 0.9 (magenta), 0.99 (red), and 0.99999 (orange). {\color{black} The corresponding analytical expressions are provided in equations \eqref{ToutTin} and \eqref{OhmacheHeat!}.} For the sake of illustration, the boundary conditions have been fixed as follows: $T_{\text{out}}/T_{\text{in}}=0.5$, and $\varrho_{\text{in}}/\varrho_{\text{out}}=0.1$. {\color{black}We observe that, at high rotation rates, $T(\varrho)$ tends to form  a local minimum near the outer boundary.}}
    \label{fig:stationary}
\end{figure}

To solve the above problem, we seek solutions of \eqref{grumpyone} or \eqref{grainofsalt} that do not depend on $t$, $\phi$, or $z$. In both equations, one is left with the differential equation
\begin{equation}
\partial_\varrho \big( \varrho \partial_\varrho T_R\big)=0 \, ,
\end{equation}
which gives $T_R=A\ln\varrho +B$. The two integration constants $A$ and $B$ can be determined from the boundary conditions at $\varrho_{\text{in}}$ and $\varrho_{\text{out}}$, giving (see figure~\ref{fig:stationary}, left panel)
\begin{equation}\label{ToutTin}
T= \dfrac{\Gamma\ln(\varrho/\varrho_{\text{in}})}{\Gamma_{\text{out}} \ln(\varrho_{\text{out}}/\varrho_{\text{in}})}T_{\text{out}} - \dfrac{\Gamma \ln(\varrho/\varrho_{\text{out}})}{\Gamma_{\text{in}}\ln(\varrho_{\text{out}}/\varrho_{\text{in}})} T_{\text{in}} \, .
\end{equation}
Then, the heat flux four-vector $q^\mu$ can be computed from \eqref{heattttone}, giving (see figure~\ref{fig:stationary}, right panel)
\begin{equation}\label{OhmacheHeat!}
\begin{bmatrix}
q^t \\
q^\varrho \\
q^\phi \\
q^z \\
\end{bmatrix}= \dfrac{T_{\text{in}}/\Gamma_{\text{in}}{-}T_{\text{out}}/\Gamma_{\text{out}}}{\ln(\varrho_{\text{out}}/\varrho_{\text{in}})} \, \dfrac{\kappa \Gamma}{\varrho} \,
\begin{bmatrix}
0\\
1\\
0\\
0\\
\end{bmatrix} \, .
\end{equation}
As one might have expected, if $T_{\text{in}}/\Gamma_{\text{in}}> T_{\text{out}}/\Gamma_{\text{out}}$, some outgoing heat flux is generated, which propagates radially. Besides the usual $1/\varrho$ decay law, $q^\varrho$ comes with an additional Lorentz factor $\Gamma(\varrho)$, which enhances the flow of heat at large $\Omega\varrho$. This relativistic correction originates from the fact that the conserved heat current \eqref{JHHH} has a Lorentz factor in the denominator, namely\footnote{Note that, in a rotating body, the heat current is not just the energy current, but rather $J_H^\mu =T^{\mu t}-\Omega T^\mu_\phi$ (see section~\ref{heatcurrentsection}), which may be viewed as the energy current ``in the rotating frame''. Algebraically, the $1/\Gamma$ suppression factor appears because, at relativistic speeds, the angular-momentum term $\Omega T^\mu_\phi$ becomes comparable to the energy term $T^{\mu t}$, and the two balance each other. Physically, the same $1/\Gamma$ factor can be interpreted as a gravitational redshift effect, as viewed from the perspective of rotating observers.}
\begin{equation}\label{JHHH2}
J_H^\mu = \dfrac{\varepsilon u^\mu +q^\mu}{\Gamma} \, . 
\end{equation}
To see the connection between these two Lorentz factors explicitly, one can integrate the conservation law $\nabla_\mu J_H^\mu =0$ over a spacetime region bounded by two radii, $\varrho_1$ and $\varrho_2$, and then invoke Gauss' theorem, which leads directly to the scaling 
$q^\varrho \varrho/\Gamma=\text{const}$, in agreement with~\eqref{OhmacheHeat!}. 

\subsection{Second application: Trend to equilibrium}
\vspace{-0.3cm}

\begin{figure}[h!]
    \centering
\includegraphics[width=0.45\linewidth]{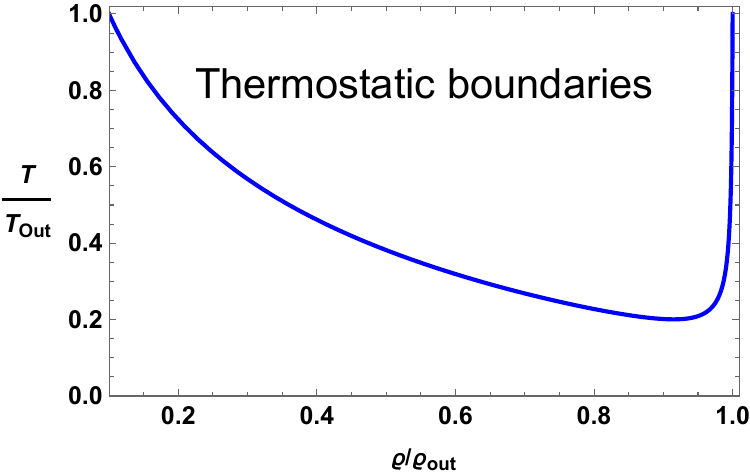}
\includegraphics[width=0.45\linewidth]{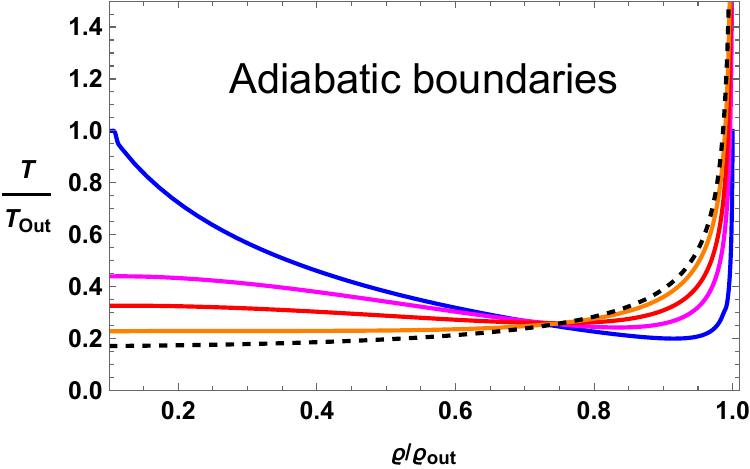}
\caption{Left panel: Stationary temperature profile {\color{black}\eqref{ToutTin}} of a cylindrical shell with $\varrho_{\text{in}}/\varrho_{\text{out}}=0.1$ and $\Omega=0.999$, in thermal contact with an inner and an outer thermostat having the same $T$. Right panel: Thermal evolution {\color{black}\eqref{TrrrEqaziamo}} of the same shell if the thermostats are quickly removed, and replaced with an adiabatic insulator. The various curves are snapshots at different times following the change in boundary conditions, namely $Dt/\varrho_{\text{out}}^2=0$ (blue), 0.05 (magenta), 0.1 (red), 0.2 (orange), $\infty$ (dashed). {\color{black} The temperature profile spontaneously evolves to the Tolman-Ehrenfest prediction \eqref{Tolmaniamo}, as expected.} }
    \label{fig:coating}
\end{figure}

Working in the same setting as in section~\ref{SectionB}, suppose that the two thermostats are set to the same temperature, i.e. $T_{\text{in}}=T_{\text{out}}$. In a non-relativistic setting, this would imply that the system is in thermodynamic equilibrium. However, this is no longer true in relativity, where $T\propto\Gamma$ in equilibrium, which would imply $T_{\text{out}}>T_{\text{in}}$. Consequently, if we externally enforce $T_{\text{in}}=T_{\text{out}}$, some heat spontaneously flows from the inner to the outer boundary of the shell, giving $q^\varrho>0$. Near the rotation axis, where the centripetal acceleration is negligible (i.e. $u^\lambda\nabla_\lambda u_\nu \approx 0$), this flux causes the temperature to decrease with increasing $\varrho$, as captured by Fourier's law: $\partial_\varrho T \approx -q^\varrho/\kappa<0$; see equation \eqref{heattttone}. Conversely, when $\Omega\varrho$ approaches 1, equation \eqref{Tolmaniamo} indicates that the temperature should increase with a factor $\Gamma$. These competing trends balance at intermediate values of $\varrho$, producing a temperature minimum, clearly visible in the left panel of figure~\ref{fig:coating}.

To verify that there is indeed a flow of energy from $\varrho_{\text{in}}$ to $\varrho_{\text{out}}$, let us see what happens if we suddenly remove the thermostats, and we instantaneously coat both surfaces with an adiabatic material. In this new setting, no energy is supplied at the boundaries, so the total thermal energy inside the shell, i.e.
\begin{equation}\label{Uthth}
U_{\text{th}}=\int_{\text{Shell}}  J_H^\mu d\Sigma_\mu = Nc_v \dfrac{\int_{\varrho_{\text{in}}}^{\varrho_{\text{out}}} T_R \Gamma \varrho d\varrho}{\int_{\varrho_{\text{in}}}^{\varrho_{\text{out}}} \Gamma \varrho d\varrho} +\mathcal{O}(\partial^2) \, ,
\end{equation}
is a conserved quantity \cite[\S 3.3.2]{PoissonToolkit2009pwt} ($N\,{=}\,\int nu^\mu d\Sigma_\mu$ is the total particle number, and $\varepsilon_{\text{eq}}\,{=}\,nc_vT$, since $nc_v\,{=}\,\text{const}$).
As a result, nothing should stop the initial flux $q^\varrho>0$ from driving heat towards the outer surface, causing a decrease in $T(\varrho_{\text{in}})$  and a corresponding increase in $T(\varrho_{\text{out}})$, until the equilibrium state with $T\propto \Gamma$ is reached as~$t\rightarrow \infty$. 

The above expectations are indeed confirmed in figure~\ref{fig:coating} (right panel), where we numerically solve the appropriate boundary-value problem for the given scenario:
\begin{equation}\label{TrrrEqaziamo}
\begin{cases}
\partial_{t} T_R=  \dfrac{D}{ \Gamma \varrho} \partial_{\varrho} \big( \varrho \partial_{\varrho} T_R\big) \, , \\
\partial_{\varrho} T_R(t,\varrho_{\text{in}})= \partial_\varrho T_R(t,\varrho_{\text{out}})=0 \, ,\\
T_R(0,\varrho)=\dfrac{\Gamma_{\text{in}} \ln(\varrho/\varrho_{\text{in}})- \Gamma_{\text{out}} \ln(\varrho/\varrho_{\text{out}})}{\Gamma_{\text{in}}\Gamma_{\text{out}} \ln(\varrho_{\text{out}}/\varrho_{\text{in}})} T_{\text{out}}   \, .\\
\end{cases} 
\end{equation}
The first line is equation \eqref{grainofsalt} under the assumption of cylindrical symmetry. The second line imposes the condition of zero heat flux across the boundaries (recall that $q^\varrho \propto -\partial_\varrho T_R$). The last line specifies the initial temperature profile \eqref{ToutTin}, with $T_{\text{out}}\,{=}\,T_{\text{in}}$. Note that, if we multiply the first line of \eqref{TrrrEqaziamo} by $\Gamma \varrho$, integrate in $\varrho$, and invoke the second line, we find that \eqref{Uthth} (removing the order $\partial^2$) is indeed a constant of motion.

The attentive reader may notice that, strictly speaking, the boundary data in \eqref{TrrrEqaziamo} is inconsistent at $(t,\varrho)=(0,\varrho_{\text{in}})$ and $(t,\varrho)=(0,\varrho_{\text{out}})$. Such inconsistency arises because, as we replace the thermostats with adiabatic coating, the heat flux at the boundaries is \textit{instantaneously} reduced from \eqref{OhmacheHeat!} to zero. To model this transition realistically, we ``flatten'' $T_R(0,\varrho)$ in a tiny neighborhood of the boundaries (see blue curve in figure~\ref{fig:coating}). The details of the function used in this flattening are practically irrelevant, as parabolic theories such as \eqref{grainofsalt} rapidly erase all small-scale structures.

\newpage
\subsection{Third application: Cooling of a hot cylinder with thermostatic boundary}
\vspace{-0.35cm}

\begin{figure}[h!]
    \centering
\includegraphics[width=0.43\linewidth]{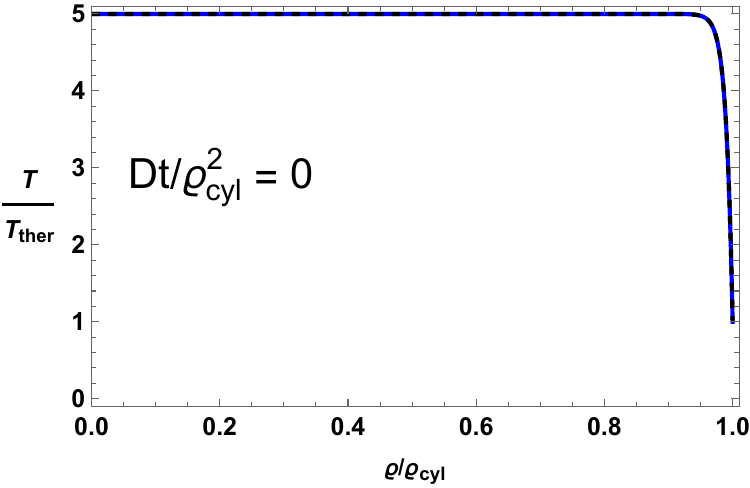}
\includegraphics[width=0.43\linewidth]{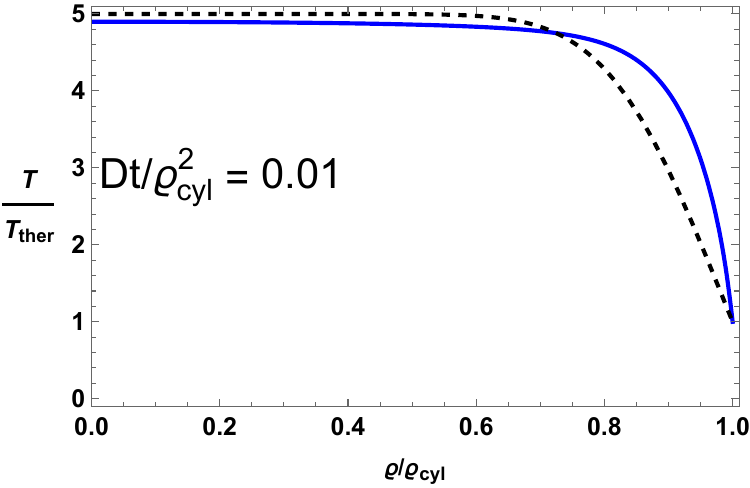}
\includegraphics[width=0.43\linewidth]{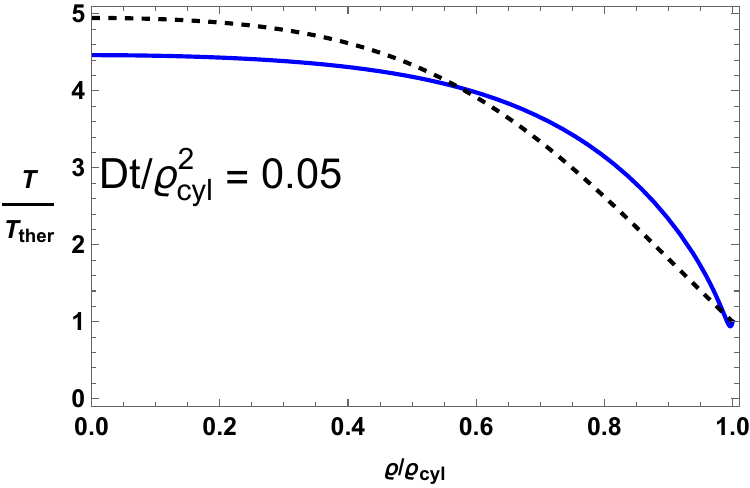}
\includegraphics[width=0.43\linewidth]{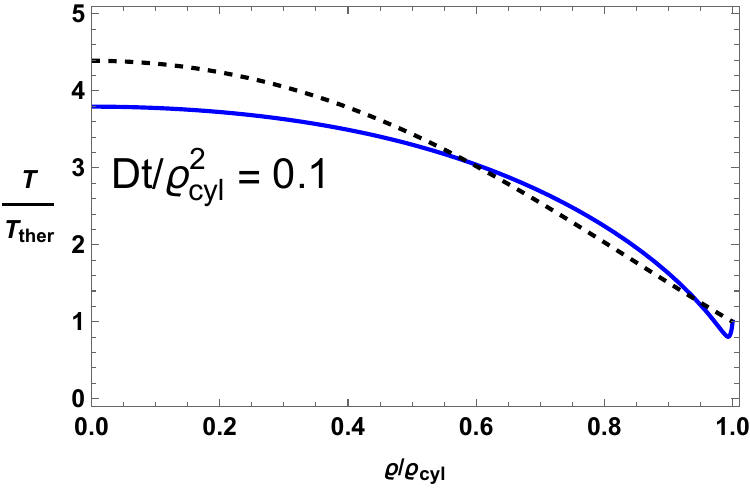}
\includegraphics[width=0.43\linewidth]{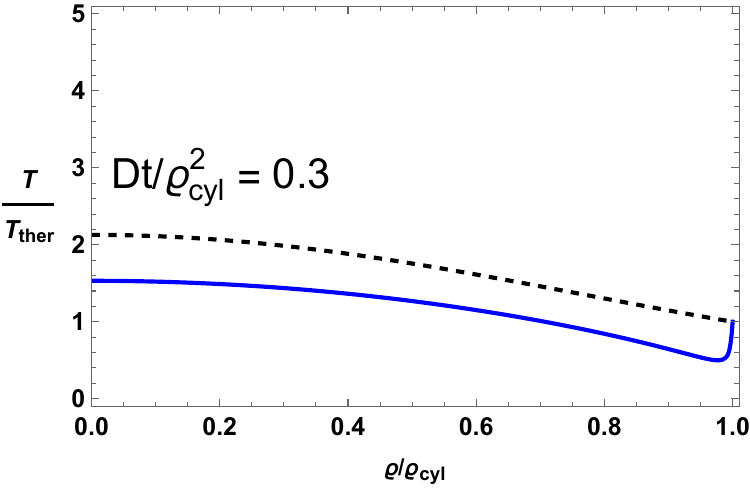}
\includegraphics[width=0.43\linewidth]{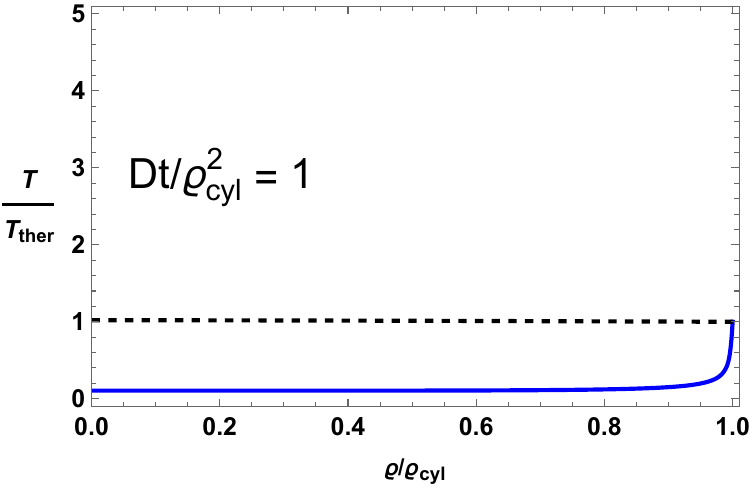}
\caption{Thermal evolution {\color{black}\eqref{Cauchyiamostrani}} of a hot infinite cylinder whose surface is in contact with a thermostat with temperature $T_{\text{ther}}$. The dashed line is the non-rotating case, while the blue line refers to a cylinder spinning with surface speed~$\Omega\varrho_{\text{cyl}}=0.999$. {\color{black}At high rotation rates, the temperature is initially driven outward by the centrifugal term ($q^\mu \sim -u^\nu \nabla_\nu u^\mu$). Subsequently, the temperature profile develops a local minimum (as in figure \ref{fig:stationary}), before relaxing to the Tolman–Ehrenfest solution \eqref{Tolmaniamo}.}}
\label{fig:EvolvingThermostat}
\end{figure}

As a last application, we consider a minimal cooling scenario: a cylinder of radius $\varrho_{\text{cyl}}$, rotating with angular velocity $\Omega {=} 0.999/\varrho_{\text{cyl}}$, starts with an almost flat temperature profile given by $T_{\text{init}}(\varrho) =(5 {-}4 \varrho^{100}/\varrho_{\text{cyl}}^{100}) T_{\text{ther}}$, and its surface is in contact with a thermostat at temperature $T_{\text{ther}}$. These conditions define the following boundary-value problem:
\begin{equation}\label{Cauchyiamostrani}
\begin{cases}
\partial_{t} T_R=  \dfrac{D}{ \Gamma \varrho} \partial_{\varrho} \big( \varrho \partial_{\varrho} T_R\big) \, , \\
\partial_{\varrho} T_R(t,0)=0 \, , \\
T_R(t,\varrho_{\text{cyl}})=T_{\text{ther}} \, ,\\
T_R(0,\varrho)=T_\text{init}(\varrho)/\Gamma(\varrho) \, ,\\
\end{cases} 
\end{equation}
where the second line imposes that the temperature be smooth on the rotation axis. The numerical solution of \eqref{Cauchyiamostrani} is shown in figure~\ref{fig:EvolvingThermostat}. As before, we see that, due to the acceleration term in the heat flux \eqref{heattttone}, the temperature drops very rapidly near the outer boundary, which results in the temporary formation of a local temperature minimum.

\newpage
\section{Cooling of rotating bodies via black-body emission}

This article primarily focuses on heat transfer via conduction. However, radiative processes play a crucial role in astrophysical contexts, as they constitute the main cooling mechanism for isolated objects in vacuum. For this reason, we devote this final section to examining how special-relativistic effects (like time dilation and length contraction) alter the cooling rate of rotating black bodies. Some of the following results may already be present in the literature, but we are unaware of a fully covariant equation derived from first principles.

\subsection{Cooling formula in curved spacetime}

Fix a Cauchy surface $\Sigma$, and construct from it a one-parameter family of subsequent hypersurfaces $\Sigma_\mathfrak{t}$, obtained by translating $\Sigma$ along the Lie flow $x^\mu \rightarrow  x^\mu(\mathfrak{t})$  generated by the Killing vector $\K^\mu{=}\K u^\mu$. Since this flow is an isometry, all hypersurfaces $\Sigma_\mathfrak{t}$ are geometrically identical (same ``shape''). Let us integrate the divergence of the bivector $\mathcal{B}^{\mu \nu}=J_H^\mu \K^\nu -J_H^\nu \K^\mu$ over the support of the rotating body within $\Sigma_{\mathfrak{t}}$. Stokes' theorem \cite[\S 3.3.3]{PoissonToolkit2009pwt} then gives
\begin{equation}
\int_{\text{Body}(\mathfrak{t})} \nabla_\nu \mathcal{B}^{\mu \nu} d^3 \Sigma_\mu =\dfrac{1}{2} \int_{\text{Surface}(\mathfrak{t})} \mathcal{B}^{\mu \nu} d^2 S_{\mu \nu} \, .
\end{equation}
Since $\nabla_\nu J_H^\nu =\nabla_\nu \K^\nu=0$, we find that $\nabla_\nu \mathcal{B}^{\mu \nu}=\K^\nu \nabla_\nu J_H^\mu -J_H^\nu \nabla_\nu \K^\mu=(\mathcal{L}_\K J_H)^\mu$, where $\mathcal{L}$ is the Lie derivative. Moreover, due to the geometric decomposition \eqref{JHHH}, we have $\mathcal{B}^{\mu \nu}=\K^2 (q^\mu u^\nu {-}q^\nu u^\mu)$, and we are left with
\begin{equation}
\int_{\text{Body}(\mathfrak{t})} (\mathcal{L}_\K J_H)^\mu \, d^3 \Sigma_\mu = \int_{\text{Surface}(\mathfrak{t})} \K^2 q^{\mu} u^{\nu}  d^2 S_{\mu \nu} \, .
\end{equation}
Now we need to plug in the explicit formulas for $J_H^\mu$ and $q^\mu$. Inside the body, the constitutive relation \eqref{BDNKone} holds, so we can insert it in the volume integral. In the surface integral, we invoke the black-body law, according to which the energy flux $q^\mu$ carried by the emitted photons has norm $\sigma T^4$ ($\sigma =\, $Stefan-Boltzmann constant), and has a direction that is normal to both $u^\mu$ and to the surface element of the medium\footnote{\label{foOwuiotz}Note that the black-body emission law $q^\mu =\sigma T^4 \times \text{``unit normal''}$ is valid only if photons emitted from one point of the surface are not reabsorbed at another point. This is valid only if the shape of the body is convex, and if we can neglect the possibility of some photon geodesics ``falling back'' onto the body due to gravity \cite{Baines:2023bgm}.}. But this is the same direction as that of the vector $u^\nu dS_{\nu}^\mu$, since $dS_{\mu \nu}$ is antisymmetric (so $u^\mu u^\nu dS_{\mu \nu}{=}0$), and is normal to the surface element by definition. Hence, the contraction $q^{\mu} u^{\nu}  d^2 S_{\mu \nu}$ just gives the product of the norms. Accounting for the orientation signs, we obtain 
\begin{equation}\label{boduzko}
\int_{\text{Body}(\mathfrak{t})}  nc_v  \mathcal{L}_\K (\K T) \,  u^\mu \,d^3 \Sigma_\mu-\int_{\text{Body}(\mathfrak{t})}  [\mathcal{L}_\K (\kappa \nabla (\K T))]^\mu \,d^3 \Sigma_\mu =- \int_{\text{Surface}(\mathfrak{t})} \sigma \K^2 T^4 || u^{\nu}  d^2 S_{\mu \nu}|| \, .
\end{equation}
Till this point, our treatment was exact. Now, let's make the usual assumption that the cooling process is quasi-static, in the sense that all changes in time $\mathfrak{t}$ are slow (i.e. $\mathcal{L}_\K \text{``fields''}\rightarrow 0$), and the Tolman-Ehrenfest law is approximately valid at all times (i.e. $\nabla_\mu (\K T)\rightarrow 0$). This allows us to neglect the second integral in \eqref{boduzko}, being of higher order in deviations from the quasi-static approximation. Furthermore, we can regard $\K T\equiv T_R(\mathfrak{t})$ as uniform across $\text{Body}(\mathfrak{t})$, and we can factor it out of the remaining integrals. Considering that $\mathcal{L}_\K(\K T)=\mathcal{L}_\K T_R(\mathfrak{t})=dT_R(\mathfrak{t})/d\mathfrak{t}$, we finally obtain an ordinary differential equation for the redshifted temperature:
\begin{equation}
\label{capolavorooooooo!!!}
\boxed{ \dfrac{dT_R(\mathfrak{t})}{d\mathfrak{t}} =- \dfrac{\int_{\text{Surface}(\mathfrak{t})} \, \K^{-2} || u^{\nu}  d^2 S_{\mu \nu}||}{\int_{\text{Body}(\mathfrak{t})}  nc_v  \,  u^\mu \,d^3 \Sigma_\mu} \, \sigma T_R(\mathfrak{t})^4 \,  .}
\end{equation}
This is the cooling law of rigidly rotating black bodies in general relativity, expressed in a manifestly covariant form. {\color{black} The main assumptions we used to arrive here are: (a) rigid motion, (b) quasistatic evolution, (c) pure surface emission, and (d) negligible self-irradiation (and the various assumptions it entails, see footnote \ref{foOwuiotz}).}

Let us note that, since all surfaces $\Sigma_{\mathfrak{t}}$ are geometrically identical, the only way the prefactor in \eqref{capolavorooooooo!!!} depends on $\mathfrak{t}$ is through $c_v$ (which is a function of the temperature). In the particular case where the specific heat scales with a power law of the temperature, $c_v=h(n)T^\ell$, we can factor out $T_R^\ell$, and integrate the differential equation, which gives
\begin{equation}
T_R(\mathfrak{t})^{\ell-3}=T_R(0)^{\ell-3} -(\ell{-}3) \dfrac{\int_{\text{Surface}} \, \K^{-2} || u^{\nu}  d^2 S_{\mu \nu}||}{\int_{\text{Body}}  nh(n)\K^{-\ell}  \,  u^\mu \,d^3 \Sigma_\mu} \, \sigma \mathfrak{t} \spc (\text{for }\ell\neq 3) \, ,
\end{equation}
or an exponential decay for $\ell=3$.

\subsection{Flat spacetime limit}
\vspace{-0.2cm}

Let us specialize equation \eqref{capolavorooooooo!!!} to a system in Minkowski spacetime, with flat metric $ds^2=-dt^2{+}dx^2{+}dy^2{+}dz^2$. As usual, we assume rigid rotation, with $u^\mu \partial_\mu =\Gamma[  \partial_t+\Omega (x \partial_y {-} y \partial_x)]$, where $\Omega=\text{const}$ and $\Gamma=[1-\Omega^2 (x^2{+}y^2)]^{-1/2}$. Since $\partial_t$ and $x \partial_y {-} y \partial_x$ are Killing vectors \cite[\S 3.8]{carroll_2019}, the motion is indeed rigid in Born's sense, with $\K=1/\Gamma$. Therefore, the relevant Killing field is $\K^\mu \partial_\mu =\partial_t+\Omega (x \partial_y {-} y \partial_x)$, which generates the flow
\begin{equation}
\begin{bmatrix}
t\\
x\\
y\\
z\\
\end{bmatrix}
\rightarrow
\begin{bmatrix}
t+\mathfrak{t}\\
x\cos(\Omega \mathfrak{t})-y\sin(\Omega \mathfrak{t})\\
y\cos(\Omega \mathfrak{t})+x\sin(\Omega\mathfrak{t})\\
z \\
\end{bmatrix} \, .
\end{equation}
We choose $\Sigma=\{t{=}0\}$ as our initial hypersurface, so $\Sigma_{\mathfrak{t}}=\{t{=}\mathfrak{t}\}$. From this, we immediately have $u^\mu d^3\Sigma_\mu =\Gamma d^3x$. The two-dimensional normal surface element can be expressed as follows:
\begin{equation}
d^2 S_{\mu \nu} =
\begin{bmatrix}
0 & \mathfrak{n}_x & \mathfrak{n}_y & \mathfrak{n}_z \\
-\mathfrak{n}_x & 0 & 0 & 0 \\
-\mathfrak{n}_y & 0 & 0 & 0 \\
-\mathfrak{n}_z & 0 & 0 & 0 \\
\end{bmatrix} d^2 S \, ,
\end{equation}
where $(\mathfrak{n}_x, \mathfrak{n}_y, \mathfrak{n}_z)$ and $d^2 S$ are the usual unit normal and surface element in 3D Euclidean space. Introducing the ordinary three-velocity $(v^x,v^y,v^z)=(-y\Omega,x\Omega,0)$, equation \eqref{capolavorooooooo!!!} becomes
\begin{equation}\label{capolavoro2}
\dfrac{dT_R(t)}{dt} =- \dfrac{\int_{\text{Surface}(t)} \, \Gamma^3 \sqrt{1{-}(v^j \mathfrak{n}_j)^2} \, d^2 S}{\int_{\text{Body}(t)}  nc_v  \,  \Gamma d^3 x} \, \sigma T_R(t)^4 \,  .
\end{equation}

\vspace{-0.2cm}
\subsection{Consistency check}
\vspace{-0.2cm}

To confirm that equation \eqref{capolavoro2} accurately captures all special-relativistic effects, we test it in a scenario where the correct answer is knowable by independent means. Consider the following setup: A small parallelepiped rotates around the origin with angular velocity $\Omega$, constrained to move in a circular path by a rope of length $\varrho$ anchored at the origin. The parallelepiped has a constant heat capacity $C$, and its rest-frame edge lengths are $L_1$, $L_2$, and $L_3$,  all negligibly short compared to $\varrho$. The rope is attached to the center of face ``12'', with edge ``2'' aligned with the direction of the tangential velocity. Assuming the rope is adiabatic and the system is placed in a vacuum, we ask: What is the law governing the radiative cooling of the parallelepiped in the laboratory frame?

To solve this problem, we don't really need equation \eqref{capolavoro2}. Given that the parallelepiped is very small, all its points instantaneously travel with the same speed $\varrho \Omega$. Thus, we can solve the cooling problem in the instantaneous rest frame of the parallelepiped, and then invoke the usual time dilation formula, with $\Gamma=(1{-}\Omega^2 \varrho^2)^{-1/2}$. This gives
\begin{equation}\label{soluviammone}
T(t)^{-3}=T(0)^{-3}+6 \,  \dfrac{L_1 L_2 {+} L_2 L_3 {+}L_3 L_1}{C\Gamma} \, \sigma t \, .  
\end{equation}
Let us verify that \eqref{capolavoro2} gives exactly the same prediction, under the assumption that $\Gamma$ is approximately uniform across the parallelepiped (the latter being very small). By definition, we have $T_R =\K T=T/\Gamma$. The quantity $nc_v$ is the heat capacity per unit volume as measured in the rest frame, which is just $C/(L_1L_2L_3)$. In the laboratory frame, the edges have lengths $L_1$, $L_2/\Gamma$, and $L_3$. Hence, equation \eqref{capolavoro2} simplifies to
\begin{equation}\label{capolavoro3}
\dfrac{dT(t)}{dt} =- \dfrac{\int_{\text{Surface}} \, \sqrt{1{-}(v^j \mathfrak{n}_j)^2} \, d^2 S}{C} \, \sigma T(t)^4 \,  .
\end{equation}
We still need to evaluate the surface integral. Opposite faces give exactly the same contribution, so that
\begin{equation}
\begin{split}
\int_{\text{Surface}} \, \sqrt{1{-}(v^j \mathfrak{n}_j)^2} \, d^2 S ={}& 2\int_{\text{Face }12} \!\! \! \!\! \! \!\! \! \!\! \!  \sqrt{1{-}(v^3)^2} \, d^2 S+2\int_{\text{Face }23} \!\! \! \!\! \! \!\! \! \!\! \! \sqrt{1{-}(v^1)^2} \, d^2 S+2\int_{\text{Face }31} \!\! \! \!\! \! \!\! \! \!\! \! \sqrt{1{-}(v^2)^2} \, d^2 S \\
={}& 2 L_1(L_2/\Gamma) +2 (L_2/\Gamma) L_3 + 2 \Gamma^{-1} L_3 L_1 =2(L_1L_2 +L_2 L_3 +L_3 L_1)/\Gamma \, .\\
\end{split}
\end{equation}
Plugging this formula into \eqref{capolavoro3}, and solving for $T(t)$, we recovered equation~\eqref{soluviammone}. 

It is useful to examine how the Lorentz factors balance out to give consistent outcomes in the two approaches. In the first (more intuitive) approach, the factor $1/\Gamma$ in \eqref{soluviammone} stems from time dilation. In the second (more formal) approach, that same factor accounts for a reduction of the emitted heat due to relativistic phenomena that depend on the orientation of the face. In particular, the two faces whose normal is parallel to the velocity have a reduced heat flux due to the time dilation, accounted for by the factor $\sqrt{1{-}(v^j \mathfrak{n}_j)^2}$, while the other four faces have the same heat flux as in the rest frame, but the total emission is reduced due to the surface area being Lorentz contracted. 

\newpage
\section{Conclusions}

A longstanding open problem in relativistic thermodynamics has been resolved: the derivation of a first-principles model for heat propagation in rotating rigid media. We have obtained a fully covariant equation of motion for the temperature, which generalizes and improves upon the non-covariant cooling models currently employed in neutron star physics. This equation is hyperbolic, causal, and stable, and it admits a well-posed initial-value formulation for initial data on arbitrary Cauchy surfaces. Moreover, we have demonstrated that any alternative theory of heat transport must reduce to our formulation at leading order in the derivative expansion. In other words, equation \eqref{BDNKHeat} cannot be further improved, unless one is willing to include additional degrees of freedom or go to higher orders in the derivative expansion. These results were only made possible by recent advances in our global understanding of relativistic fluid dynamics\footnote{\color{black} Our approach closely parallels that of BDNK theory \cite{BemficaDNDefinitivo2020,Kovtun2019}, as it is formulated within a first-order derivative expansion in which the definition of temperature is fixed \textit{a posteriori} to ensure causality, stability, and well-posedness. Extending the expansion to second order would lead to a Burnett-type theory for heat conduction, analogous in spirit to (non-resummed) BRSSS \cite{Baier2008}, but formulated in a causal hydrodynamic frame. In contrast, our framework departs substantially from transient hydrodynamic theories (e.g. Cattaneo's theory \cite{cattaneo1958}, Israel–Stewart theory \cite{Israel_Stewart_1979,Hishcock1983,Denicol2012Boltzmann}, divergence-type theories \cite{Liu1986,GerochLindblom1990}, and Carter's multifluid theory \cite{lopez2011,carter1991,PriouCOMPAR1991}), which treat the heat flux $q^\mu$ as an independent dynamical variable. Nonetheless, when such theories are truncated to first order in derivatives, they are all known to reproduce equation \eqref{eckartone}, which is equivalent to \eqref{BDNKHeat} up to a field redefinition.} \cite{BemficaDNDefinitivo2020,Kovtun2019,GavassinoSuperlum2021,Basar:2024qxd,GavassinoParabolic2025hwz}.

In addition to having solid theoretical foundations, our framework is as flexible as equation \eqref{newtonian}. It accommodates a wide range of boundary conditions, depending on whether the surface is in contact with a thermostat, an adiabatic insulator, or a vacuum (see, e.g., figure~\ref{fig:coating}). Moreover, since the right-hand side of \eqref{BDNKHeat} involves a standard wave operator, expressing all terms explicitly in a chosen coordinate system is relatively straightforward, unlike cases involving the projector $\Delta^{\mu \nu}$, which tends to complicate the expressions. 

{\color{black} As an astrophysical application, equation \eqref{BDNKHeat} can be used to describe the cooling of self-gravitating, rapidly rotating compact objects (such as millisecond-period neutron stars) provided that the spacetime metric remains approximately unaffected by temperature changes (otherwise the existence of a timelike Killing vector field would be invalidated). This ``fixed-metric approximation'' is not a severe simplification, as it is already standard in conventional cooling models. To go beyond this approximation, one would need to develop a full viscoelastic theory of matter incorporating heat conduction, in which the evolution equation for $T$ is coupled to those for $n$ and $u^\mu$. In that case, a simple heat-conduction equation such as \eqref{BDNKHeat} would no longer suffice to model the evolution.}

\section*{Acknowledgements}

This work is partially supported by a Vanderbilt Seeding Success Grant, the IN2P3 Master Project NewMAC and the ANR Project GW-HNS (ANR-22-CE31-0001-01).
 
\appendix

\section{Derivation of the Tolman-Ehrenfest effect}\label{AppAAA}

There are several ways to demonstrate that, in thermodynamic equilibrium, a macroscopic medium possesses a collective velocity $u^\mu$ and a temperature $T$ such that the ratio $u^\mu/T$ forms a Killing vector field \cite{TolmanLaw1930,BecattiniBeta2016,Santiago:2019aem}. The easiest one consists of verifying that this relation indeed holds for the velocity and temperature of the local black-body radiation spectrum generated by the given medium. We shall follow such an approach below.

Let $S_{\text{tot}}=S_M+S_P$ and $Q_{\text{tot}}^I=Q_M^I+Q_P^I$ be the entropy and the conserved charges (energy, momentum...) of an isolated ``matter+radiation'' system, where quantities with surbscript ``$M$'' refer to the medium, and quantities with subscript ``$P$'' refer to the photon gas. In equilibrium, the total system must be in a state that maximizes $S_{\text{tot}}$ for fixed values of $Q_{\text{tot}}^I$. Therefore, defined some Lagrange multipliers $\alpha_I$, we require that
\begin{equation}
\delta (S_M+S_P)+\alpha_I \delta (Q_M^I+Q_P^I) =0 \, ,
\end{equation}
for all possible variations within the physical state space of the system. Now, the macroscopic states of the matter component are characterized by a list of material degrees of freedom $\Psi$, while the kinetic states of the photon gas are characterized by the invariant distribution function $f(x^\mu,p^\alpha)$, which counts how many photons occupy a photon eigenmode located at the event $x^\mu$ and with four-momentum $p^\alpha$. Since $\Psi$ and $f$ are independent variables, we have
\begin{equation}\label{variaitons}
\begin{split}
\dfrac{\delta S_M}{\delta \Psi} +\alpha_I \dfrac{\delta Q_M^I}{\delta \Psi} ={}& 0 \, , \\ 
\dfrac{\delta S_P}{\delta f} +\alpha_I \dfrac{\delta Q_P^I}{\delta f} ={}& 0 \, , \\ 
\end{split}
\end{equation}
which tells us that, in equilibrium, the medium and the photons have the same intensive parameters $\alpha_I =-\partial S/\partial Q^I$. Now we only need to determine $f$. To this end, we recall that, according to kinetic theory \cite{cercignani_book,mihalas_book}, all extensive variables can be expressed as integrals in the single-photon phase space:
\begin{equation}\label{phasespace}
\begin{split}
S_P ={}& \int_{\text{Phase space}} \big[{-}f \ln f+(1{+}f)\ln (1{+}f)\big] \,\, d \, \text{``Phase volume''} \, , \\
Q_P^I ={}& \int_{\text{Phase space}} f q^I \,\, d \, \text{``Phase volume''} \, , \\
\end{split}
\end{equation}
where $q^I(x^\mu,p^\alpha)$ is the quantum of conserved charge carried by a photon located at the given point in phase space. Plugging \eqref{phasespace} into the second line \eqref{variaitons}, we obtain, after some algebra, $f\,{=}\,(e^{-\alpha_I q^I}{-}1)^{-1}$, which has a Bose-Einstein form. But since the photon is its own antiparticle, the only conserved quantum numbers it can carry are energy, momentum, and angular momentum, namely all charges of the form $q^I\,{=}\,{-} \K^I_\alpha p^\alpha$, where $\K^I$ are Killing vector fields. Therefore, we have that $\alpha_I q^I=-\alpha_I\K^I_\alpha p^\alpha=\beta_\alpha p^\alpha$, where we have introduced the new field $\beta_\alpha=-\alpha_I\K^I_\alpha$, which is also a Killing vector (since the multipliers $\alpha_I$ are constants). Hence, we finally arrive at
\begin{equation}
f=\dfrac{1}{e^{-\beta_\alpha p^\alpha}-1} \, .
\end{equation}
This is a standard black-body distribution \cite{UdeyIsrael1982}, since we can always rewrite 
\begin{equation}\label{babibube}
-\beta_\alpha p^\alpha =\dfrac{\text{``Energy''}}{\text{``Temperature''}}= \dfrac{-u_\alpha p^\alpha}{T} \, ,
\end{equation}
where $u^\alpha$ is the local velocity of the black body (i.e. the velocity of the observer who measures an isotropic spectrum) and $T$ is its temperature (as viewed by that same observer). Comparing the sides of \eqref{babibube}, we conclude that $u^\mu/T=\beta^\mu$, where the latter is a Killing vector field. This concludes our proof.

\bibliography{Biblio}

\label{lastpage}

\end{document}